\documentclass{ifacconf}

\usepackage{graphicx,xcolor,amsmath,siunitx,amssymb,mathtools}      
\usepackage{booktabs,tikz,pgfplots,siunitx}    
\usepackage{natbib}        
\definecolor{myred}{rgb}{0.6350, 0.0780, 0.1840}
\definecolor{mygreen}{rgb}{0.4660, 0.6740, 0.1880}
\definecolor{myblue}{rgb}{0, 0.4470, 0.7410}
\begin{document}
\begin{frontmatter}

\title{Modeling and Predictive Control for the Treatment of Hyperthyroidism\thanksref{footnoteinfo}} 

\thanks[footnoteinfo]{This project has received funding from the European Research Council (ERC) under the European Union’s Horizon 2020 research and innovation programme (grant agreement No 948679). 
}

\author[IRT]{Tobias M. Wolff} 
\author[IRT]{Maylin Menzel} 
\author[Bochum,Blankenstein,Witten]{Johannes W. Dietrich}
\author[IRT]{Matthias A. Müller}

\address[IRT]{Leibniz University Hannover, Institute of Automatic Control, Germany, (e-mail: \{wolff, mueller\}@irt.uni-hannover.de).}
\address[Bochum]{Diabetes, Endocrinology and Metabolism Section, Department of Internal Medicine~I, St. Josef Hospital, Ruhr University Bochum, Germany, (e-mail: johannes.dietrich@ruhr-uni-bochum.de).}
\address[Blankenstein]{Diabetes Centre Bochum-Hattingen, St. Elisabeth-Hospital Blankenstein, Hattingen, Germany.}
\address[Witten]{Ruhr Center for RareDiseases (CeSER), Ruhr University of Bochum and Witten/Herdecke University, Bochum, Germany.}

\begin{abstract}                
In this work, we propose an approach to determine the dosages of antithyroid agents to treat hyperthyroid patients. Instead of relying on a trial-and-error approach as it is commonly done in clinical practice, we suggest to determine the dosages by means of a model predictive control (MPC) scheme. To this end, we first extend a mathematical model of the pituitary-thyroid feedback loop such that the intake of methimazole, a common antithyroid agent, can be considered. Second, based on the extended model, we develop an MPC scheme to determine suitable dosages. In numerical simulations, we consider scenarios in which (i) patients are affected by Graves' disease and take the medication orally and (ii) patients suffering from a life-threatening thyrotoxicosis, in which the medication is usually given intravenously. Our conceptual study suggests that determining the medication dosages by means of an MPC scheme could be a promising alternative to the currently applied trial-and-error approach. 
\end{abstract}


\end{frontmatter}

\section{Introduction}
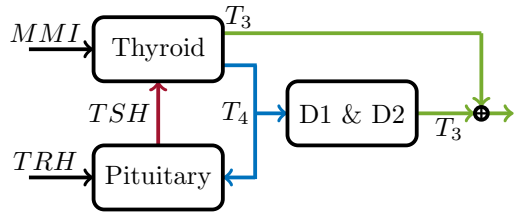
\begin{figure}[t!]
	\centering
	\vspace{5pt}
	\begin{tikzpicture}[scale=0.85]
		\draw[line width = .5mm,->](0,0) -- (1,0);
		\node[align=center] at (0.25,.25) {$MMI$};
		\draw[line width = .5mm, color=mygreen](3,0.25) -- (7,0.25);
		\draw[line width = .5mm, color=mygreen,->](7,0.25) -- (7,-0.9);
		\draw[line width = .5mm, color=myblue](3,-0.25) -- (3.525,-0.25);
		\draw[line width = .5mm, color=myblue,->](3.5,-1) -- (4,-1);
		\draw[line width = .5mm,->, color=myred](2,-1.5) -- (2,-0.5);
		\draw[line width = .5mm, rounded corners] (1,-.5) rectangle (3,.5);
		\node[align=center] at (2,0) {Thyroid};
		\node[align=center] at (3.25,.5) {$T_3$};
		\draw[line width = .5mm,->, color=mygreen](6,-1) -- (6.9,-1);
		\draw[line width = .5mm, rounded corners] (4,-1.5) rectangle (6,-.5);
		\node[align=center] at (5,-1) {D1 \& D2};
		\node[align=center] at (6.5,-1.25) {$T_3$};
		\draw[line width = .5mm,->, color=mygreen](7.1,-1) -- (7.5,-1);
		\draw[line width = .5mm] (7,-1) circle (0.1);
		\draw[line width = .3mm](6.9,-1) -- (7.1,-1);
		\draw[line width = .3mm](7,-0.9) -- (7,-1.1);
		\node[align=center] at (3.2,-1) {$T_4$};
		\draw[line width = .5mm, color=myblue](3.5,-0.25) -- (3.5,-2.025);
		\draw[line width = .5mm,->, color=myblue](3.5,-2) -- (3,-2);
		\draw[line width = .5mm, rounded corners] (1,-2.5) rectangle (3,-1.5);
		\node[align=center] at (2,-2) {Pituitary};
		\node[align=center] at (1.4,-1) {$TSH$};
		\draw[line width = .5mm,->](0,-2) -- (1,-2);
		\node[align=center] at (0.25,-1.75) {$TRH$};
	\end{tikzpicture}
	\caption{\small Simplified block diagram of the pituitary-thyroid feedback loop. The thyroid produces the hormones $T_3$ and $T_4$. On the one hand, this process is stimulated by $TSH$, a hormone secreted from the pituitary. On the other hand, this process is inhibited by Methimazole (MMI), an antithyroid agent (i.e., a medication), which constitutes the control input in this work. The hormone $T_4$ gets converted into $T_3$ by the enzymes 5'-deiodinase type I (D1) and 5'-deiodinase type II (D2). Finally, $TRH$ (which is a natural hormone, and not a medication) stimulates the production of $TSH$.}
	\label{fig:HPT_axis}
\end{figure}

The pituitary-thyroid feedback loop is a natural control loop in the human body. As illustrated in Fig.~\ref{fig:HPT_axis}, the main mechanisms are the following \citep{gardner2000basic}. The pituitary secretes thyroid-stimulating-hormone ($TSH$), which stimulates at the thyroid gland the synthesis of the hormones triiodothyronine ($T_3$) and thyroxine ($T_4$). In peripheral organs like the liver and the kidney, $T_4$ is converted into the biologically active $T_3$ by 5'-deiodinase type I (D1) and 5'-deiodinase type~II (D2). Moreover, in the pituitary, central $T_3$ inhibits the production of $TSH$, whereas thyrotropin-releasing hormone ($TRH$) stimulates the synthesis of $TSH$.

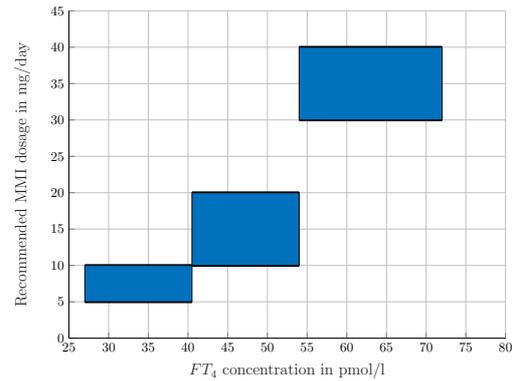
\begin{figure}[t!]
	\centering
	\scalebox{0.5}{
%
%
\definecolor{mycolor1}{rgb}{0.00000,0.44700,0.74100}%
\begin{tikzpicture}

\begin{axis}[%
width=4.521in,
height=3.418in,
at={(0.758in,0.628in)},
scale only axis,
xmin=25,
xmax=80,
xlabel style={font=\color{white!15!black}},
xlabel={\large $FT_4$ concentration in pmol/l},
ymin=0,
ymax=45,
ylabel style={font=\color{white!15!black}},
ylabel={\large Recommended MMI dosage in mg/day},
axis background/.style={fill=white},
axis x line*=bottom,
axis y line*=left,
xmajorgrids,
ymajorgrids
]
\addplot [color=black, line width=2.0pt, forget plot]
  table[row sep=crcr]{%
27	5\\
27.5	5\\
28	5\\
28.5	5\\
29	5\\
29.5	5\\
30	5\\
30.5	5\\
31	5\\
31.5	5\\
32	5\\
32.5	5\\
33	5\\
33.5	5\\
34	5\\
34.5	5\\
35	5\\
35.5	5\\
36	5\\
36.5	5\\
37	5\\
37.5	5\\
38	5\\
38.5	5\\
39	5\\
39.5	5\\
40	5\\
40.5	5\\
};
\addplot [color=black, line width=2.0pt, forget plot]
  table[row sep=crcr]{%
27	10\\
27.5	10\\
28	10\\
28.5	10\\
29	10\\
29.5	10\\
30	10\\
30.5	10\\
31	10\\
31.5	10\\
32	10\\
32.5	10\\
33	10\\
33.5	10\\
34	10\\
34.5	10\\
35	10\\
35.5	10\\
36	10\\
36.5	10\\
37	10\\
37.5	10\\
38	10\\
38.5	10\\
39	10\\
39.5	10\\
40	10\\
40.5	10\\
};

\addplot[area legend, draw=black, fill=mycolor1, fill opacity=0.3, forget plot]
table[row sep=crcr] {%
x	y\\
27	5\\
27.5	5\\
28	5\\
28.5	5\\
29	5\\
29.5	5\\
30	5\\
30.5	5\\
31	5\\
31.5	5\\
32	5\\
32.5	5\\
33	5\\
33.5	5\\
34	5\\
34.5	5\\
35	5\\
35.5	5\\
36	5\\
36.5	5\\
37	5\\
37.5	5\\
38	5\\
38.5	5\\
39	5\\
39.5	5\\
40	5\\
40.5	5\\
40.5	10\\
40	10\\
39.5	10\\
39	10\\
38.5	10\\
38	10\\
37.5	10\\
37	10\\
36.5	10\\
36	10\\
35.5	10\\
35	10\\
34.5	10\\
34	10\\
33.5	10\\
33	10\\
32.5	10\\
32	10\\
31.5	10\\
31	10\\
30.5	10\\
30	10\\
29.5	10\\
29	10\\
28.5	10\\
28	10\\
27.5	10\\
27	10\\
}--cycle;
\addplot [color=black, line width=2.0pt, forget plot]
  table[row sep=crcr]{%
40.5	10\\
41	10\\
41.5	10\\
42	10\\
42.5	10\\
43	10\\
43.5	10\\
44	10\\
44.5	10\\
45	10\\
45.5	10\\
46	10\\
46.5	10\\
47	10\\
47.5	10\\
48	10\\
48.5	10\\
49	10\\
49.5	10\\
50	10\\
50.5	10\\
51	10\\
51.5	10\\
52	10\\
52.5	10\\
53	10\\
53.5	10\\
54	10\\
};
\addplot [color=black, line width=2.0pt, forget plot]
  table[row sep=crcr]{%
40.5	20\\
41	20\\
41.5	20\\
42	20\\
42.5	20\\
43	20\\
43.5	20\\
44	20\\
44.5	20\\
45	20\\
45.5	20\\
46	20\\
46.5	20\\
47	20\\
47.5	20\\
48	20\\
48.5	20\\
49	20\\
49.5	20\\
50	20\\
50.5	20\\
51	20\\
51.5	20\\
52	20\\
52.5	20\\
53	20\\
53.5	20\\
54	20\\
};

\addplot[area legend, draw=black, fill=mycolor1, fill opacity=0.3, forget plot]
table[row sep=crcr] {%
x	y\\
40.5	10\\
41	10\\
41.5	10\\
42	10\\
42.5	10\\
43	10\\
43.5	10\\
44	10\\
44.5	10\\
45	10\\
45.5	10\\
46	10\\
46.5	10\\
47	10\\
47.5	10\\
48	10\\
48.5	10\\
49	10\\
49.5	10\\
50	10\\
50.5	10\\
51	10\\
51.5	10\\
52	10\\
52.5	10\\
53	10\\
53.5	10\\
54	10\\
54	20\\
53.5	20\\
53	20\\
52.5	20\\
52	20\\
51.5	20\\
51	20\\
50.5	20\\
50	20\\
49.5	20\\
49	20\\
48.5	20\\
48	20\\
47.5	20\\
47	20\\
46.5	20\\
46	20\\
45.5	20\\
45	20\\
44.5	20\\
44	20\\
43.5	20\\
43	20\\
42.5	20\\
42	20\\
41.5	20\\
41	20\\
40.5	20\\
}--cycle;
\addplot [color=black, line width=2.0pt, forget plot]
  table[row sep=crcr]{%
54	30\\
54.5	30\\
55	30\\
55.5	30\\
56	30\\
56.5	30\\
57	30\\
57.5	30\\
58	30\\
58.5	30\\
59	30\\
59.5	30\\
60	30\\
60.5	30\\
61	30\\
61.5	30\\
62	30\\
62.5	30\\
63	30\\
63.5	30\\
64	30\\
64.5	30\\
65	30\\
65.5	30\\
66	30\\
66.5	30\\
67	30\\
67.5	30\\
68	30\\
68.5	30\\
69	30\\
69.5	30\\
70	30\\
70.5	30\\
71	30\\
71.5	30\\
72	30\\
};
\addplot [color=black, line width=2.0pt, forget plot]
  table[row sep=crcr]{%
54	40\\
54.5	40\\
55	40\\
55.5	40\\
56	40\\
56.5	40\\
57	40\\
57.5	40\\
58	40\\
58.5	40\\
59	40\\
59.5	40\\
60	40\\
60.5	40\\
61	40\\
61.5	40\\
62	40\\
62.5	40\\
63	40\\
63.5	40\\
64	40\\
64.5	40\\
65	40\\
65.5	40\\
66	40\\
66.5	40\\
67	40\\
67.5	40\\
68	40\\
68.5	40\\
69	40\\
69.5	40\\
70	40\\
70.5	40\\
71	40\\
71.5	40\\
72	40\\
};

\addplot[area legend, draw=black, fill=mycolor1, fill opacity=0.3, forget plot]
table[row sep=crcr] {%
x	y\\
54	30\\
54.5	30\\
55	30\\
55.5	30\\
56	30\\
56.5	30\\
57	30\\
57.5	30\\
58	30\\
58.5	30\\
59	30\\
59.5	30\\
60	30\\
60.5	30\\
61	30\\
61.5	30\\
62	30\\
62.5	30\\
63	30\\
63.5	30\\
64	30\\
64.5	30\\
65	30\\
65.5	30\\
66	30\\
66.5	30\\
67	30\\
67.5	30\\
68	30\\
68.5	30\\
69	30\\
69.5	30\\
70	30\\
70.5	30\\
71	30\\
71.5	30\\
72	30\\
72	40\\
71.5	40\\
71	40\\
70.5	40\\
70	40\\
69.5	40\\
69	40\\
68.5	40\\
68	40\\
67.5	40\\
67	40\\
66.5	40\\
66	40\\
65.5	40\\
65	40\\
64.5	40\\
64	40\\
63.5	40\\
63	40\\
62.5	40\\
62	40\\
61.5	40\\
61	40\\
60.5	40\\
60	40\\
59.5	40\\
59	40\\
58.5	40\\
58	40\\
57.5	40\\
57	40\\
56.5	40\\
56	40\\
55.5	40\\
55	40\\
54.5	40\\
54	40\\
}--cycle;
\end{axis}
\end{tikzpicture}%
}
	\caption{\small Illustration of the guidelines given by the American Thyroid association (ATA) \citep{Ross2016}. The treating physician may encounter many difficulties to select one precise dosage using these (rather vague) recommendations. }
\label{fig:guidelines}
\end{figure}

One of the most important diseases related to the pituitary-thyroid feedback loop is Graves' disease, in which autoimmune antibodies (more specifically thyrotropin receptor antibodies) overstimulate the production of thyroid hormones. Consequently, the hormone concentrations of $T_3$ and $T_4$ increase and the concentration of $TSH$ declines - a condition called hyperthyroidism. Several options exist to treat hyperthyroidism: First, one can surgically remove (parts of) the thyroid gland, called thyroidectomy, which leads almost always to hypothyroidism (denoting a lack of thyroid hormones) and goes along with the usual risks of a surgery. Second, one can eliminate thyroid tissue with radioactive iodine, which, on the one hand, exposes patients to radioactive radiation and, on the other hand, also leads in many cases to hypothyroidism. Third, one can inhibit the (over-)production of thyroid hormones by antithyroid agents such as methimazole (MMI) (sometimes also referred to as thiamazole). This option is usually the first choice of physicians treating hyperthyroidism in Europe, since it does not go along with lifelong hypothyroidism as thyroidectomy and does not expose the body to radioactive radiation \citep{Ross2016}. Therefore, a success of the approach based on antithyroid agents is crucial. However, finding accurate dosages for patients is challenging. There currently exist only rough guidelines for physicians on how to determine the dosages, see Fig.~\ref{fig:guidelines}, which is based on the guidelines developed in \cite{Ross2016} (compare also a similar illustration in \cite{benninger2023hyperthyroidism}). For instance, consider as a special case a patient that has a free $T_4$ ($FT_4$) concentration\footnote{Free $T_4$ and free $T_3$ ($FT_3$) denote the hormone concentrations that are not bound to specific binding proteins.} of 54 pmol/l. In that case, the treating physician needs to choose a dosage between 10 mg/day up to 40 mg/day, which constitutes a very large interval. Furthermore, the prescription typically goes along in a trial-and-error fashion, meaning that the physician prescribes an initial dosage and evaluates whether this dosage normalizes the hormone concentrations and reduces the symptoms. If this is not the case, the physician adapts the dosage (possibly several times). On the one hand, if the prescribed dosage is too low, the patient might continuously suffer from hyperthyroid symptoms. On the other hand, if the prescribed dosage is inappropriately high, the patient might suffer from symptoms of hypothyroidism and unnecessary side effects.  

An appealing alternative to improve this suboptimal trial-and-error approach is a prescription of MMI using model-based control. \cite{TheilerSchwetz2022} propose to determine the dosages of MMI by means of a simple (and hence easy to use) mathematical model of the pituitary-thyroid feedback loop and a proportional integral controller. However, \cite{TheilerSchwetz2022} do not consider an oral intake of antithyroid agents in the mathematical model. 

In this paper, we extend a mathematical model of the pituitary-thyroid feedback loop developed in \cite{Dietrich_Dok2001,Berberich2018,Wolff2022AHDS} to allow an intravenous and an oral intake of antithyroid agents. Furthermore, we design a model predictive control (MPC) scheme to determine optimal dosages in case of (i) patients affected by Graves' disease, which take in MMI orally once a day and (ii) patients suffering from a life-threatening thyrotoxicosis (denoting a substantial excess of thyroid hormones that is associated with weight loss, atrial fibrillation, and, rarely, death \citep{Ross2016}), in which MMI is given intravenously twice (or three) times per day. 

This approach leads to the following contributions. Our simulation results suggest that the MPC scheme based on the here extended mathematical model could be a good alternative to replace the current trial-and-error approach in the two mentioned scenarios. Furthermore, our results give additional evidence to the current treatment guidelines of prescribing further medications (than just MMI) in case of a life-threatening thyrotoxicosis. 

\section{Model extensions \& controller design}
\label{sec:extensions}
In this section, we explain the extensions of the mathematical model of the pituitary-thyroid feedback loop. Subsequently, we develop an MPC scheme to determine optimal MMI dosages to treat hyperthyroidism.

\subsection{Model extension}
\label{subsec:model:extension}
The differential equations of the entire mathematical model describing the pituitary-thyroid feedback loop of a generic individual and the related numerical parameter values can be found in the supplementary material of this work\footnote{The supplementary material is given here: https://doi.org/10.25835/aknnlyz7}. The mathematical model was originally developed in \cite{Dietrich_Dok2001} and further refined in \cite{Berberich2018,Wolff2022AHDS}. The reader is referred to these references for a detailed introduction to the model. In the present work, we extend this model such that the main mechanism of MMI can be considered, namely the intrathyroidal inhibition of the activity of thyroid peroxidase (TPO), which is a crucial enzyme for the synthesis of $T_3$ and $T_4$ in the thyroid gland. Precisely, TPO catalyzes the iodination (denoting an incorporation of iodine atoms into a molecule) of tyrosyl residues on thyroglobulin which is a precursor protein of the thyroid hormones, compare~\cite{Manna2013} for a detailed explanation of TPO's mechanism.

To model this mechanism, we first consider the relationship between MMI that is taken in orally once at some time $t_0$ and its subsequent plasma concentration by means of a Bateman function (compare, e.g., \cite{Garrett1994}) 
\begin{equation}
	MMI_{\mathrm{Pl}}(t) = \frac{f_{\mathrm{b}}u(t_0)k_a}{V(k_a - k_e)} \big( e^{-k_e(t - t_0)} - e^{-k_a(t-t_0)} \big),
	\label{eq:MMI_oral}
\end{equation} 
where $MMI_{\mathrm{Pl}}(t)$ denotes the concentration of MMI in the plasma at time $t$, $f_{\mathrm{b}} \in [0,1]$ the bioavailability of MMI, $u(t_0)$ the dosage of MMI at time $t_0$, $V \in \mathbb{R}_{\geq 0}$ the volume of distribution, $k_e \in \mathbb{R}_{\geq 0}$ the elimination constant, and $k_a \in \mathbb{R}_{\geq 0}$ the absorption constant. We consider a bioavailability of $f_{\mathrm{b}} = 0.93$ \citep{Jansson1985}. Next, we determine the parameters $k_a$ and $k_e$ by using the relations $k_e= \log(2)/t_{1/2}$ and $t_{\max} = \log(k_a/k_e) / (k_a - k_e)$ with the mean elimination half-life~$t_{1/2}$ and the mean peak concentration $t_{\max}$ reported in \cite{Melander1980}, which results in $k_a = 1.0196 \: 1/\si{\hour}$ and $k_e = 0.1857\: 1/\si{\hour}$, respectively. Finally, we determine the value of $V$, by evaluating the Bateman function at $t_{\max}$ (using the mean $t_{\max}$ and the mean $MMI_{\mathrm{Pl}}(t_{\max})$ as provided by \cite{Melander1980}), leading to $V=~\SI{28}{l}$. Note that within this work, we do not aim for a model that describes individual patient dynamics, but for a generic model that captures the main mechanisms of the medication intake. Since the reported individual values of $t_{\max}$ and $t_{1/2}$ vary \citep{Melander1980}, we do not expect to obtain a model that is perfectly in line with some individual patient dynamics. Nevertheless, this procedure can be used to obtain conceptual insights in the dynamics of the medication intake. Finally, the dosage $u(t_0)$ is determined by the controller, compare Section~\ref{subsec:MPC} below. 

Second, we model the transition of the plasma methimazole concentration $MMI_{\mathrm{Pl}}$ and the concentration of methimazole in the thyroid gland $MMI_{\mathrm{th}}$ by a Michaelis-Menten kinetic (which is a common procedure to model enzyme/substrate reactions \cite{Murray2002}). It results in the following differential equation
\begin{align}
	&\frac{dMMI_{\mathrm{th}}}{dt}(t) = \nonumber \\ &\hspace{0.75cm}\alpha_{\mathrm{M,th}}G_{\mathrm{M,th}}\frac{MMI_{\mathrm{Pl}}(t)}{K_{\mathrm{M,th}} + MMI_{\mathrm{Pl}}(t)} - \beta_{\mathrm{M,th}} MMI_{\mathrm{th}}
	\label{eq:MMIth}
\end{align}
with $\alpha_{\mathrm{M,th}}$, $\beta_{\mathrm{M,th}} \in \mathbb{R}_{\geq 0}$ being the dilution factor and the clearance exponent, respectively. The parameters $G_{\mathrm{M,th}} \in \mathbb{R}_{\geq 0}$ and $K_{\mathrm{M,th}}\in \mathbb{R}_{\geq 0}$ stand for the maximal transport rate and the related Michaelis-Menten constant, respectively. We choose $\alpha_{\mathrm{M,th}} = \SI{250}{1/l}$ corresponding to a volume of distribution of 4 ml as for $T_4$ in the thyroid \citep{Wolff2022AHDS}. Typically, the clearance exponent is determined by using the relationship $\beta_{\mathrm{M,th}} = \log(2)/t_{1/2}^{\mathrm{MMI}_{\mathrm{th}}}$ \citep{Dietrich_Dok2001}, where $t_{1/2}^{\mathrm{MMI}_{\mathrm{th}}}$ denotes the half-life of MMI in the thyroid gland. To the best of our knowledge, the intrathyroidal half-life of MMI has not yet been determined experimentally. Note that measuring intrathyroidal concentrations is involved, since this requires a thyroidectomy. However, in this work, we estimate it based on the data from \cite{mandarin1996}. In this study, patients were divided into four groups according to the amount of MMI they were prescribed before undergoing a thyroidectomy. The intrathyroidal MMI concentration was measured after the thyroidectomy. In each group, approximately half of the patients received their last dosage two hours before the thyroidectomy and the other half obtained their last dosage 16 or 26 hours before the thyroidectomy. Based on these experiments, the authors argue that the intrathyroidal half-life is longer than the plasma half-life which is within 3 - 6 hours according to \cite{Okuno1987}. Based on the results given in Tab. 1 of \cite{mandarin1996}, one can estimate that the intrathyroidal half-life must be smaller\footnote{This holds, since the plasma MMI concentration of the groups which took their last dosage two hours before the surgery is much larger compared to the plasma concentration of the groups which took their last dosage 16 (or 26) hours before the surgery.} than approximately 53 (group A), 91 (group B), 48 (group C), 84 hours (group D). Obviously, under these circumstances, we can only determine a rough estimate of the intrathyroidal MMI half-life. Here, we use $t_{1/2}^{\mathrm{MMI}_{\mathrm{th}}} = \SI{30}{\hour}$ as first approach. Note that a different value of the half-life would not affect the qualitative results (although affecting the quantitative ones) as we observed in our simulations. As previously mentioned, the focus of this work is to obtain a conceptual understanding of the working mechanism and potential of model-based MMI dosage determination. Thus, such a rough estimate is sufficient. The last step is to determine $G_{\mathrm{M,th}}$ and $K_{\mathrm{M,th}}$ by using the relations between the intrathyroidal hormone concentrations and the daily MMI dosage given in \cite[Fig. 2]{mandarin1996}. The numerical values of these parameters can be found in the supplementary material of this work (see footnote 2).

Third, we establish a relation between the intrathyroidal MMI concentration and the inhibition of the activity of TPO. To this end, we exploit the results of \cite{Taurog1976}, which are based on a model incubation system. As illustrated by the different curves in Fig.~9 of \cite{Taurog1976}, this relation depends on the concentration of inorganic iodide. The concentration of inorganic iodide in the thyroid is approximately $10^{-3}\: \si{\mole}/\mathrm{l}$ \citep{Schaffhauser2005}. Then, we model the relation between the intrathyroidal MMI concentration and the activity of TPO by the following function
\begin{equation}
	TPO_{\mathrm{a}}(t) = c_0 \big( 1+ \exp(-c_1(-MMI_{\mathrm{th}}(t)^{1/c_2} +c_3))\big)^{-1}.
	\label{eq:relation_iodine}
\end{equation}
This function is (heuristically) chosen since it can well represent this relation and the parameters $c_0$, $c_1$, $c_2$, $c_3 \in \mathbb{R}_{\geq 0}$ are estimated by fitting (\ref{eq:relation_iodine}) to the curve from Fig.~9 of \cite{Taurog1976} representing an iodide concentration of $10^{-3}\: \si{\mole} /\mathrm{l}$ in a least-squares fashion. The determined numerical parameter values can be found in Table~1 of the supplementary material (see footnote 2).

Finally, we need to incorporate the modelled effects of MMI into the overall model of the pituitary-thyroid feedback loop developed in \cite{Dietrich_Dok2001,Berberich2018,Wolff2022AHDS}. In the overall model, the parameter $G_T$ (appearing in the differential equation of $T_{\mathrm{4, th}}$) represents the secretory capacity of the thyroid gland and can be interpreted as main parameter describing the thyroid hormone production rate. Since this production rate now depends on the medication intake, we adapt this previously constant parameter as follows
\begin{align}
	G_T(t) = G_{\mathrm{T,nom}} G_{\mathrm{T,co}}TPO_{\mathrm{a}}(t),
	\label{eq:G_T}
\end{align}
where $G_{\mathrm{T,nom}} \in \mathbb{R}_{\geq 0}$ stands for the nominal (euthyroid) thyroid hormone production rate, $G_{\mathrm{T,co}} \in \mathbb{R}_{\geq 0}$ for the coefficient to consider different degrees of severity of hyperthyroidism, and with $TPO_{\mathrm{a}}(t)$ resulting from \eqref{eq:relation_iodine}.

\subsection{MPC scheme}
\label{subsec:MPC}
Based on these model extensions, we implement an MPC scheme \citep{Rawlings2020}. The general principle of MPC is the following. At each sampling instant $t = k\delta, \: k \in \mathbb{N}_0$, we measure the system's state $x(t)$ and solve an optimal control problem to find the optimal input (i.e., MMI dosage) sequence with respect to a cost function over some control horizon $T$, which is an integer multiple of the sampling time $\delta$. Next, we apply the first $\ell \in \mathbb{N}_{\geq 1}$ elements of the optimal input sequence to the system. In standard MPC, $\ell = 1$, i.e., only the first element of the input sequence is applied to the systems. Here, we choose $\ell = T$ for the treatment of Graves' disease and $\ell = 3$ days for the treatment of the thyrotoxicosis. This choice is due to the fact that new hormone concentration measurements are typically not taken too frequently in outpatient treatments. Finally, the system's state is measured again and the whole process is repeated. The here applied setting for nonlinear systems in continuous time is as follows: At each sampling instant~$t$, solve
\begin{equation}
	\min_{\bar{u}_t} J\big(x(t), \bar{u}_t\big)  \label{eq:MPC}
\end{equation}
with
\begin{align}
	J(x(t),\bar{u}_t)  
	=&  \int_{t}^{t+T} ||\bar{x}(\tau; t) - x_s||_Q^2 d\tau \nonumber \\
	+& \sum_{k=0}^{T/ \delta-1} ||\bar{u}(t+k\delta;t)||_{R_1}^2 \nonumber \\
	&+ ||\bar{u}(t+k\delta;t) -\bar{u}^\ast(t-\delta; t -\ell \delta)||_{R_2}^2, \label{eq:cost_functional}
\end{align}
subject to the constraints 
\begin{align*}
	&\dot{\bar{x}}(\tau;t) = f(\bar{x}(\tau;t), \bar{u}(\tau;t)), \quad \bar{x}(t;t) = x(t), \quad \bar{u}_t \in \mathcal{U}^{T/\delta} 
\end{align*}
for $\tau \in [t, t + T]$ and $\bar{u}_t := \{\bar{u}(t;t),\bar{u}(t+\delta;t), \dots, \bar{u}(t+T-\delta;t)\}$, which is the optimization variable. Note that we only optimize over the input $\bar{u}$ at the discrete time points $t+k\delta$, since the medication is typically taken in orally, i.e., at discrete time points. The input $\bar{u}$ for all time instants is given by 
\begin{equation}
	\bar{u}(\tau;t):= \sum_{k=0}^{T/\delta-1}\bar{u}(t+k\delta;t)\delta_{\mathrm{Di	}}(\tau-(t+k\delta)), 
\end{equation}
which means that the input is zero except for the sampling instants $t + k\delta$, as described by the Dirac-delta impulse $\delta_{\mathrm{Di}}$. In (\ref{eq:cost_functional}), $J$ denotes the cost function to be minimized. The notations $\bar{x}(\tau; t)$ and $\bar{u}(\tau;t)$ stand for the predicted state~$\bar{x}$ and the predicted input~$\bar{u}$ at time $\tau$, predicted at time~$t$, respectively, and $\bar{u}^*(\cdot;t)$ and $\bar{x}^*(\cdot;t)$ denote the optimal predicted trajectories. The setpoint $x_s$ corresponds to the steady-state hormone concentrations of healthy individuals. The closed-loop input, i.e., the input that is applied to the system, is defined as $u(\tau) \coloneqq \bar{u}^\ast (\tau; t)$ for all $\tau \in [t, t+\ell\delta)$. In the cost function, besides the standard input penalty with weight $R_1$, we include a term with weight $R_2$ that penalizes the difference of the predicted input trajectory to the previously optimal input $u(t-\delta)=\bar{u}^*(t-\delta;t-\ell\delta)$ in order to keep the same dosage as long as possible to avoid frequent dosage adaptations for the patients. The weighing matrix $Q$ penalizes the difference between the predicted $T_4$, $T_3$, and $TSH$ concentrations to the setpoint, which corresponds to the healthy steady-state concentrations. The system dynamics $f(x,u)$ correspond to the right-hand sides of equations (A.1) - (A.8) (given in the supplementary material) and the newly introduced equations (\ref{eq:MMI_oral}) - (\ref{eq:G_T}). The input constraint set is defined as
\begin{align}
	&\mathcal{U} \coloneqq \{u \in \mathbb{R}| \: 0 \leq u \leq u_{\mathrm{max}} \},
\end{align}
since the input, must be limited to avoid the prescription of inadequately high dosages. Precisely, $u_{\mathrm{max}}$ is set to $\SI{35}{\milli \gram}$ in Subsection~\ref{subsec:graves:disease} and to $\SI{40}{\milli \gram}$ in Subsection~\ref{subsec:thyrotoxicosis} (since in case of a thyrotoxicosis, higher maximal dosages are allowed to bring the hormone concentration into a normal range as fast as possible). The control horizon $T$ is set to 14 days. Furthermore, the sampling time $\delta$ is chosen as $24\: \si{\hour}$ and $12 \: \si{\hour}$ in Subsection~\ref{subsec:graves:disease} (Graves' diseases) and in Subsection~\ref{subsec:thyrotoxicosis} (thyrotoxicosis), respectively. Finally, the cost function's weighting matrices $Q$, $R_1$, and $R_2$ are chosen as $Q = \mathrm{diag}(0, 10^3, 10^3, 0, 10^3, 0, 0), R_1 = 10^{-2}, R_2 = 10^{-2}$, where $\mathrm{diag}(q_1, \dots, q_n)$ denotes a diagonal matrix with elements $q_1, \dots, q_n$ on the diagonal. 
\begin{figure}[t!]
	\centering
	\includegraphics[scale=0.5]{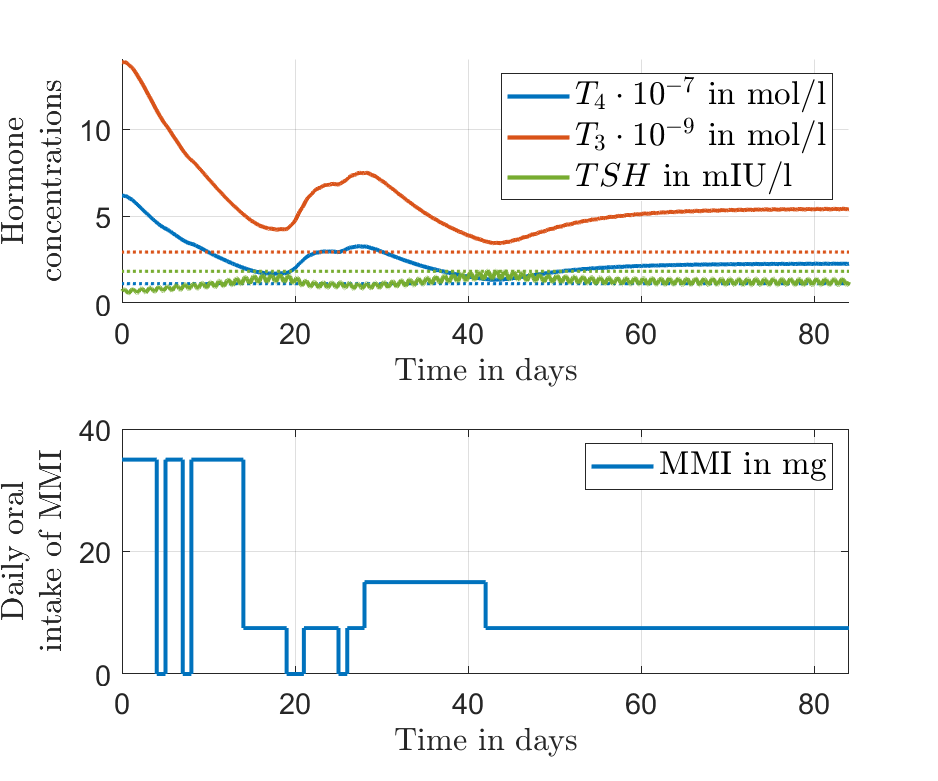}
	\caption{\small Simulation results of an exemplary treatment based on the guidelines introduced in \cite{Ross2016}. We apply the mean medication dosage for each interval, e.g., if the concentration of $FT_4$ (which is proportional to $T_4$) is within $27 - 42 \: \si{\pico\mole}/\si{\liter}$, we apply $7.5 \: \si{\milli \gram}$. As disturbances, we consider forgotten dosages and measurement noise as explained in Section~\ref{sec:simulations}.}
	\label{fig:graves:guidelines}
\end{figure}
The MPC formulation (\ref{eq:MPC}) is non-standard since the input is not applied continuously, but only at discrete time instants (once or twice per day). Moreover, the inputs applied at previous sampling instants influence the dynamics at the current and future sampling instants since the medication of, e.g., the previous day is not completely absorbed and still affects the production of thyroid hormones at the current day and the future days. When implementing the MPC scheme, instead of (\ref{eq:MMI_oral}), which describes a single medication intake, at prediction time $t+k\delta$, we consider
\begin{align}
	&MMI_{Pl}(t+k\delta;t) = \frac{fk_a}{V(k_a - k_e)}\nonumber \\
	&\times \Bigg(\sum_{\ell=0}^{t/\delta - 1} u(\ell)(e^{-k_e(t+(k-\ell)\delta)}-e^{-k_a(t+(k-\ell)\delta)}) \nonumber \\ 
	&+ \sum_{i=t/\delta}^{t/\delta + k-1} \bar{u}(i\delta;t)(e^{-k_e(t + (k-i) \delta )}-e^{-k_a(t + (k-i) \delta )})\Bigg), 
\end{align}
which denotes the predicted concentration of MMI in the plasma at time $t + k\delta$, predicted at time $t$. We need to consider (i) the medication that has already been taken in, but has not been completely absorbed (first sum), and (ii) the medication that shall be taken over the horizon, where the different dosages are considered as optimization variables (second sum). The MPC scheme is implemented\footnote{The code of all the simulations files is freely available online:  https://doi.org/10.25835/aknnlyz7.} in CasADi \citep{Andersson2019} applying the single-shooting method and using the \textit{cvodes} integrator from the SUNDIALS suite \citep{Hindmarsh2005}. 

\section{Simulation Results \& Discussion}
\label{sec:simulations}
In this section, we implement the MPC scheme based on the extended model as introduced in the previous section. On the one hand, we consider a patient affected by Graves' disease (leading to hyperthyroidism). In this case, the medication is taken orally and measurements are performed every 14 days. On the other hand, we consider patients affected by a thyrotoxicosis, where the medication is typically given intravenously and the hormone concentrations are measured every three days. 

We assume that we can measure all hormone concentrations instantaneously. Although one can measure the concentrations of $FT_4$ , $FT_3$ (which are proportional to the state $T_4$ and $T_3$) and $TSH$, the other three states ($T_{\mathrm{4, th}}$, $T_{3,c}$, and $TSH_c$) denote hormone concentrations within organs, which cannot be measured. An interesting subject for future research is the development of a state estimator such that this assumption is no longer needed. 
\begin{figure}[t!]
	\centering
	\includegraphics[scale=0.47]{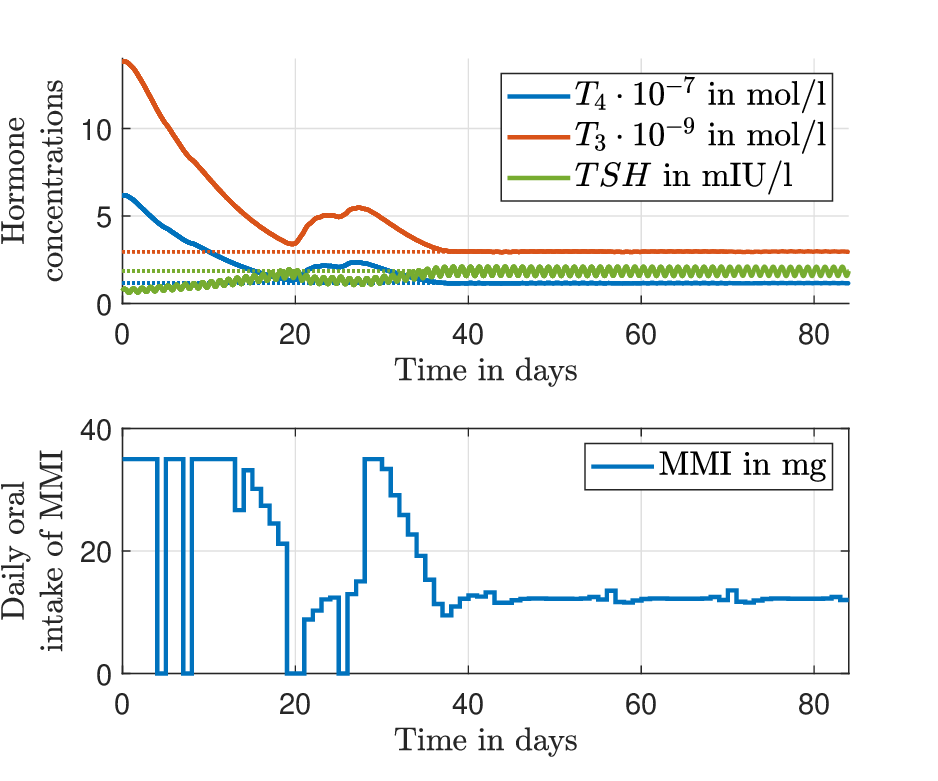}
	\caption{\small Course of the hormone concentrations and the corresponding MMI dosages that are taken orally once per day. The dosages are determined by the MPC scheme introduced in Subsection~\ref{subsec:MPC}. As explained in Section~\ref{sec:simulations}, we consider forgotten dosages, a model-plant mismatch, and some measurement noise as disturbances.}
	\label{fig:graves:MPC}
\end{figure}
\subsection{Graves' disease}
\label{subsec:graves:disease}
We first consider Graves' disease. To account for the increased thyroidal production of $T_4$ and $T_3$, we set $G_{\mathrm{T,nom}} = 1.1$, representing a 10 \% increase of the thyroidal production rate. On the one hand, we simulate a treatment based on the guidelines given in \cite{Ross2016} (compare Fig.~\ref{fig:guidelines}). On the other hand, we consider a treatment based on the MPC scheme introduced in Subsection~\ref{subsec:MPC}. We consider several disturbances. First, we consider that the (virtual) patient forgets to take their daily dosage on days 5, 8, 20, 21, 26. Second, we consider some normally distributed measurement noise using a variance which corresponds to one percent of the steady-state value of the hormone concentration. Third, we consider a model-plant mismatch in the MPC scheme. We increase the values of the parameter $G_{D1}$ and $G_{T3}$ for the model that corresponds to the true dynamics of the patient by 10 \%. Obviously, we do not increase the values of the parameters for the model used for the prediction in the MPC scheme.

\subsubsection{Treatment based on medical guidelines}
\label{subsubsec:state:of:the:art:graves}
The simulation results for the treatment based on the medical guidelines \citep{Ross2016} are given in Fig.~\ref{fig:graves:guidelines}. One can see that the (strict) application of the guidelines yields an unsatisfactory result. All hormone concentrations are substantially above (or below) their setpoints
motivating the investigation of other treatment approaches as the here suggested MPC-based approach.

\subsubsection{Treatment based on the MPC scheme}
\label{subsubsub:mpc:scheme:graves}
The simulation results for the treatment based on the MPC scheme are illustrated in Fig.~\ref{fig:graves:MPC}. The MPC scheme suggests dosages that clearly stabilize the setpoints, despite the various considered disturbances. One disadvantage of the MPC-based approach are the frequent dosage adaptations especially between days 30 - 40. These might be inconvenient for the patient and not all of the determined dosages might be available on the pharmaceutical market. This can be alleviated by (i) approximating the dosage determined by MPC with a prescription schedule that is piecewise constant (i.e., constant for several days) or (ii) by applying mixed integer-MPC schemes to only allow for discrete dosages. This would lead to a trade-off between optimal dosage and convenience for the patient.

\subsection{Thyrotoxicosis}
\label{subsec:thyrotoxicosis}
\begin{figure}[t!]
	\centering
	\includegraphics[scale=0.47]{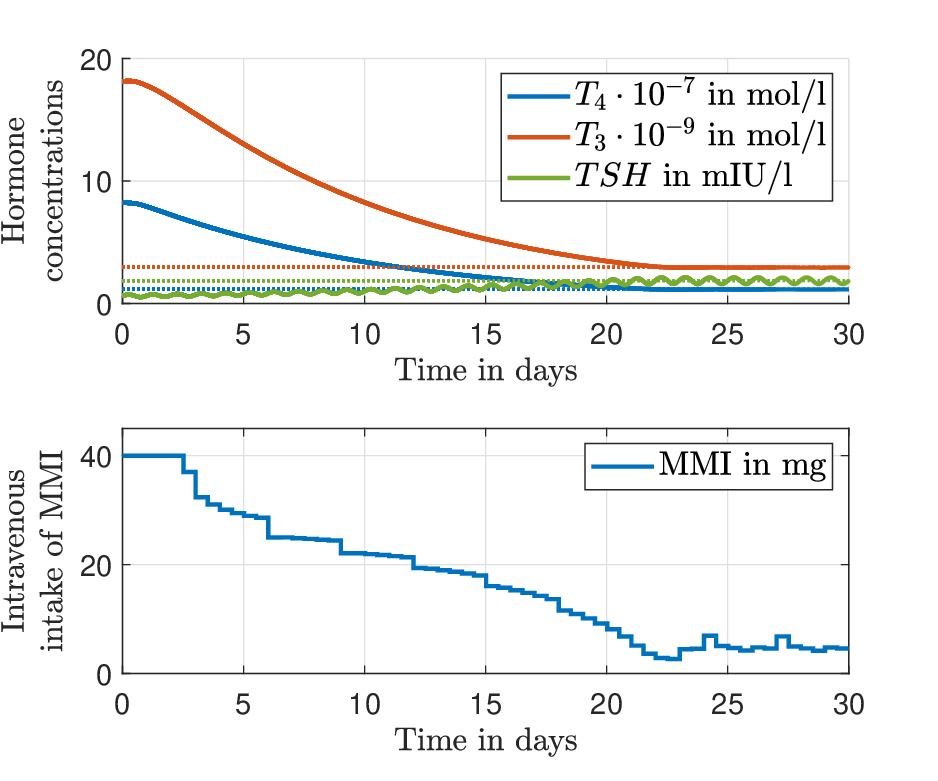}
	\caption{\small Course of the hormone concentrations and the corresponding MMI dosages that are injected intravenously twice a day. The dosages are determined by the MPC scheme introduced in Subsection~\ref{subsec:MPC}. As explained in Section~\ref{sec:simulations}, we consider a model-plant mismatch and some measurement noise as disturbances.}
	\label{fig:thyrotoxicosis:MPC}
\end{figure}
Here, we consider a (virtual) patient affected by a thyrotoxicosis. These patients are typically treated in the hospital. To represent this condition, we set $G_{\mathrm{T, nom}} = 1.15$, denoting a 15 \% increase. We here consider an intravenous infusion of MMI which is also done in clinical practice, meaning that the plasma concentration of MMI for one medication intake at time $t = t_0$ is given by
\begin{align}
	MMI_{\mathrm{Pl}}(t) = \frac{u(t_0)}{V}e^{-k_e(t-t_0)}.
\end{align}
Furthermore, we consider two medication intakes per day. Since the patient is in a critically-ill condition, we assume that the hormone concentrations are measured every 3 days. We do not consider forgotten dosages, since the medication intake is supervised by medical professionals. In turn, we still consider a model-plant mismatch and the same type of measurement noise as in the previous section. For space reasons, we only focus on the MPC-based treatment.

\subsubsection{MPC scheme}
\label{subsubsub:mpc:scheme:thyrotoxicosis}
The simulation results are illustrated in Fig.~\ref{fig:thyrotoxicosis:MPC}. It takes around 20 days to normalize the hormone concentrations, although two MMI dosages are injected per day. This is due to the long half-life of $T_4$ (7 days). Hence, even if the MMI dosages entirely inhibit the production of thyroid hormones, the normalization of the hormone concentrations is time-consuming. In general, the results underline the importance to prescribe further medication to normalize the hormone concentrations
as fast as possible. This includes medication that inhibits the release of thyroid hormones such as, e.g., lithium and/or medication that inhibits the thyroid hormones’ enterohepatic circulation to reduce their half-lives such as, e.g., colestyramine \citep{Dietrich2012}.

\section{Conclusion}
\label{sec:conclusion}
This paper presents an approach to treat hyperthyroidism using model-based predictive control. We model the mechanisms of one common antithyroid agent, namely MMI, and incorporate it into the complete model of the pituitary-thyroid feedback loop. Moreover, we design an MPC scheme to determine optimal MMI dosages for hyperthyroid patients. We illustrate the performance of the MPC scheme in case of (i) an ordinary treatment of Graves' disease and (ii) a life-threatening thyrotoxicosis. 

The obtained results suggest that the prescription of MMI based on the developed MPC framework is a promising alternative compared to the standard trial-and-error approach. While the goal of this work was to conceptually study whether and how such a model-based control framework can be used for the treatment of hyperthyroidism, different challenges need to be addressed in future work in order to be able to translate the results to clinical practice. In particular, a nonlinear observer needs to be implemented since full-state measurements are in general not available. Such an observer needs to cope with infrequent and irregular hormone concentration measurements that are typically only available in practice. To this end, we plan to build on our recent results studying sample-based detectability properties \citep{Krauss2023}. Finally, as discussed previously, while in this work we considered a generic patient model, those model parameters that are patient-specific need to be adapted to allow for individualized treatment.

\section{Acknowledgment}
We wish to thank Wolfgang Schechinger for the translation of the work by \cite{mandarin1996}.
{\footnotesize
\bibliography{ifacconf}}             
\end{document}